# Active Visual Semantics: A large-scale MEG and eye-tracking dataset for understanding visual intelligence in action

Philip Sulewski[1,2,3,+], Carmen Amme[1], Peter König[1,6], Martin N. Hebart[2,4,5,a], Tim C. Kietzmann[1,a,+]

1 Institute of Cognitive Science, Osnabrück University, 49090 Osnabrück, Germany
2 Vision and Computational Cognition Group, Max Planck Institute for Human Cognitive and Brain Sciences, 04103 Leipzig, Germany
3 Max Planck School of Cognition, Leipzig, Germany
4 Department of Medicine, Justus Liebig University Giessen, 35392 Giessen, Germany
5 Center for Mind, Brain and Behavior, Universities of Marburg, Giessen, and Darmstadt, 35032 Marburg, Germany
6 Department of Neurophysiology and Pathophysiology, Center of Experimental Medicine, University Medical Center Hamburg-Eppendorf, 20251 Hamburg, Germany
a These authors contributed equally.

+ Corresponding authors: P. Sulewski (phsulewski@gmail.com), T. C. Kietzmann (tim.kietzmann@uni-osnabrueck.de)

## Abstract

Here we present the Active Visual Semantics (AVS) dataset, a large-scale collection of magnetoencephalography (MEG) and eye-tracking data recorded while five participants freely explored 4,080 natural scenes over 10 sessions each, yielding more than 200,000 fixation epochs in total. Unlike existing neuroimaging datasets that rely on passive viewing with enforced central fixation, AVS captures brain activity during active scene exploration, including self-generated saccades and fixations. A semantic captioning task on 25% of the trials provides behavioural measures linking gaze to scene understanding and memory. In addition to neural and behavioural data, AVS includes per-fixation object category labels, human-rated annotations of the appearance of fixation targets in the scene captioning task and pupil dynamics. Individual head stabilisation casts were used during MEG data collection, which alongside with structural MRI scans, enabled precise cross-session source reconstruction. Using artificial neural network (ANN) encoding models we demonstrate that individual fixation-aligned MEG epochs hold visual content-specific signal, despite the challenges that active scene viewing poses for MEG signal quality. Further, we use fixation-aligned representation similarity analysis (RSA) and demonstrate that we can derive fixation object category averages that yield representational geometries which are highly reliable across participants. Both in MEG sensor and source space this structure is validated by its robust alignment with ANN object-level representational geometry. Taken together, AVS provides a rich resource for investigating a large variety of questions regarding the neural mechanisms of active vision, object recognition and scene captioning during natural viewing, and the relationship between gaze behaviour and memory encoding.

## Introduction

Natural vision is an inherently active process. Rather than passively receiving visual information, we continuously sample our environment through sequences of eye movements, making approximately 200,000 saccades per day. Each fixation reflects a decision about where to look and how long to remain before redirecting gaze elsewhere. While classic paradigms of vision, which closely control stimulus presentation parameters, were able to provide foundational insights into visual processing, they cannot study vision in its natural form, which includes, among others, self-guided action, target selection, information integration, memory encoding, peripheral preview, prediction, and the overall timing of vision.

One influential response to these limitations has been the development of large-scale neuroimaging datasets built around naturalistic stimuli (Allen et al., 2022; Gifford et al., 2022; Hebart et al., 2023; Zhang et al., 2025). These have proven transformative, greatly broadening the diversity and ecological validity of the images shown to participants. Yet they naturalise the stimulus while leaving the mode of sampling untouched: viewing is typically constrained by enforced fixation, omitting the self-generated, internally timed structure of visual sampling that characterises everyday perception, and with it the target selection, integration, prediction, and timing set out above.

To fill this gap, we introduce the Active Visual Semantics dataset. AVS comprises MEG and eye tracking data from five participants who freely explored 4,080 natural scenes (subsampled from the Natural Scenes Dataset (NSD) stimulus set; Allen et al., 2022) across 10 recording sessions each, yielding more than 200,000 fixation epochs in total. Stimuli were sampled to ensure uniform semantic coverage across 60 content clusters. The experimental design includes a semantic captioning task on 25% of trials, linking gaze behaviour to scene understanding and memory. The dataset provides per-fixation object category labels derived from MS-COCO and COCO-Stuff segmentations (Caesar et al., 2018; Lin et al., 2014), human-rated annotations of whether each fixation target was subsequently mentioned in the participant's scene caption, active-vision pupil dynamics, and structural MRI scans. Individualised head stabilisers were manufactured to hold the head position stable across sessions. This head position precision together with the structural scans supports reliable source reconstruction at the single-participant level. To demonstrate the dataset's utility through per-fixation encoding, we report fixation-target content from single MEG epochs, time-resolved and source-localised representational similarity analysis with noise-ceiling estimation, and behavioural validation of caption quality against the original scene annotations.

**A** Neural dynamics of active vision for natural scene understanding

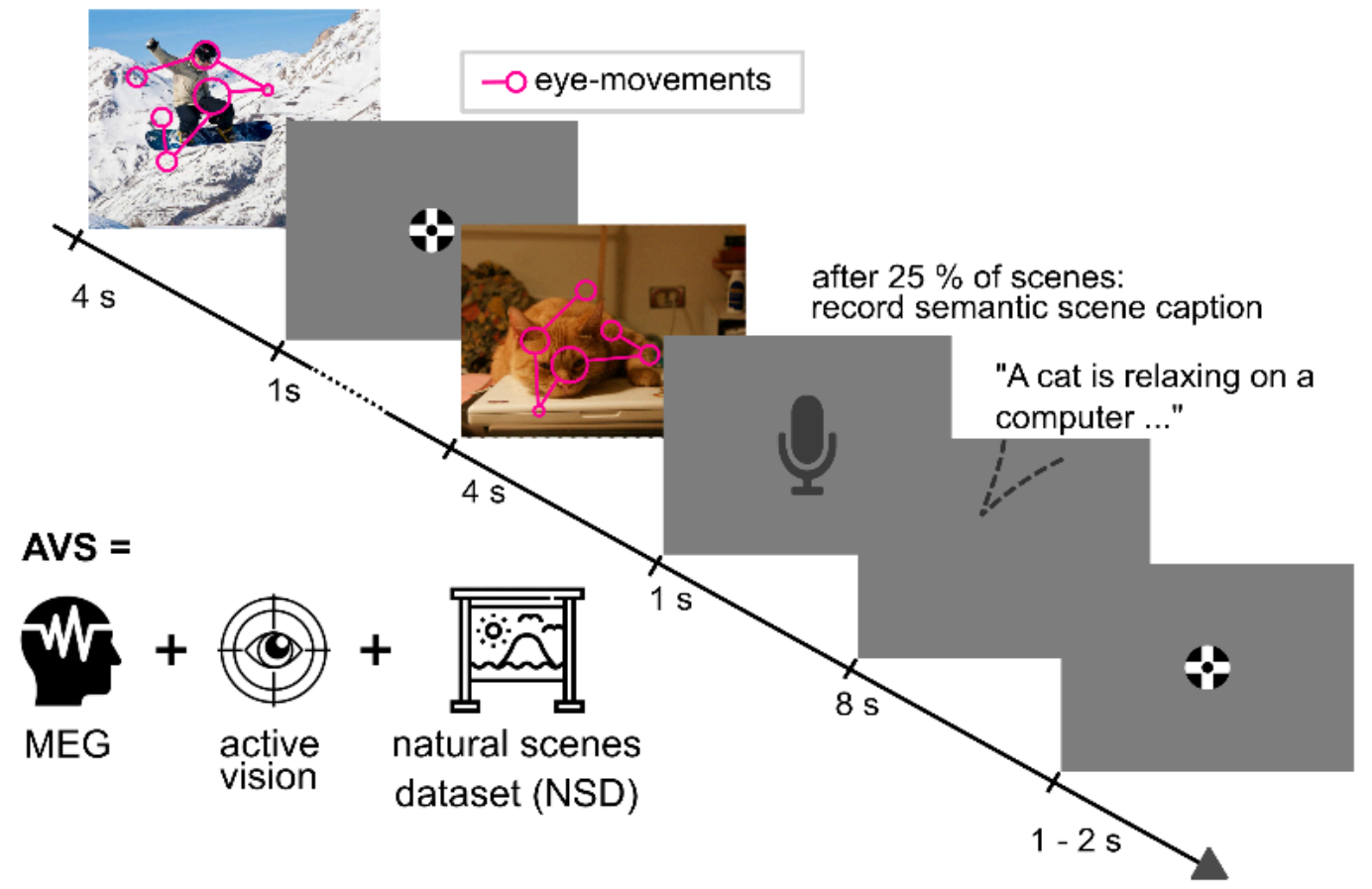

**B**

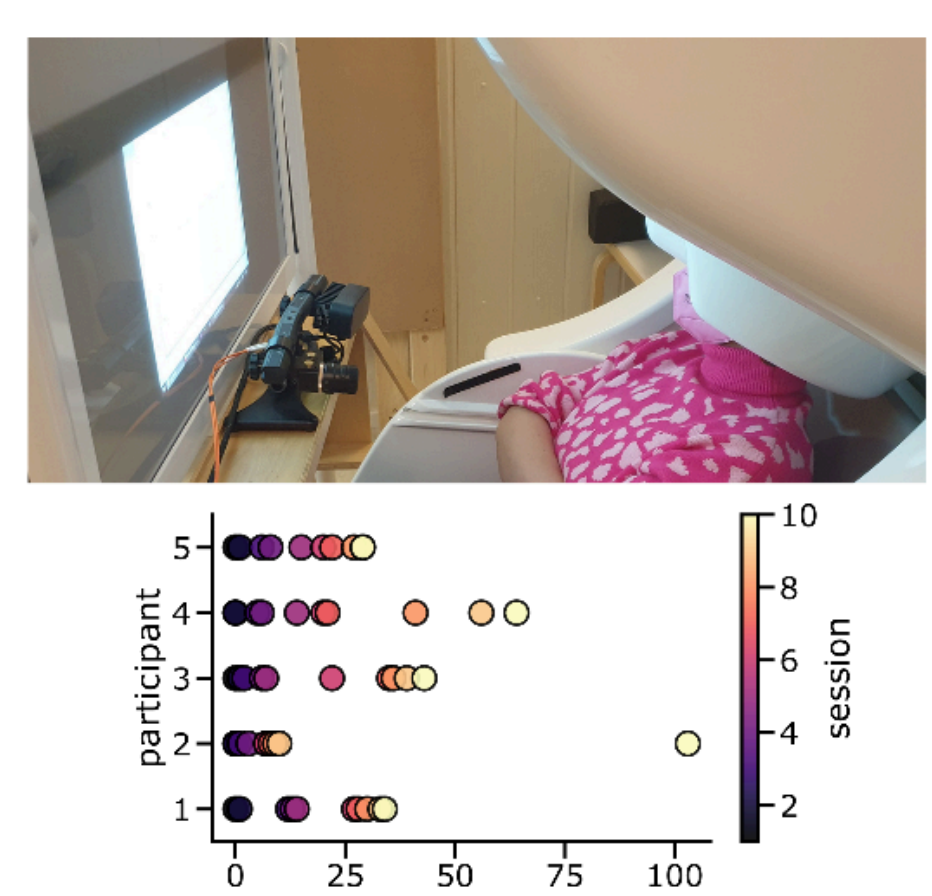

**C** Precision measurement with head stabilisers

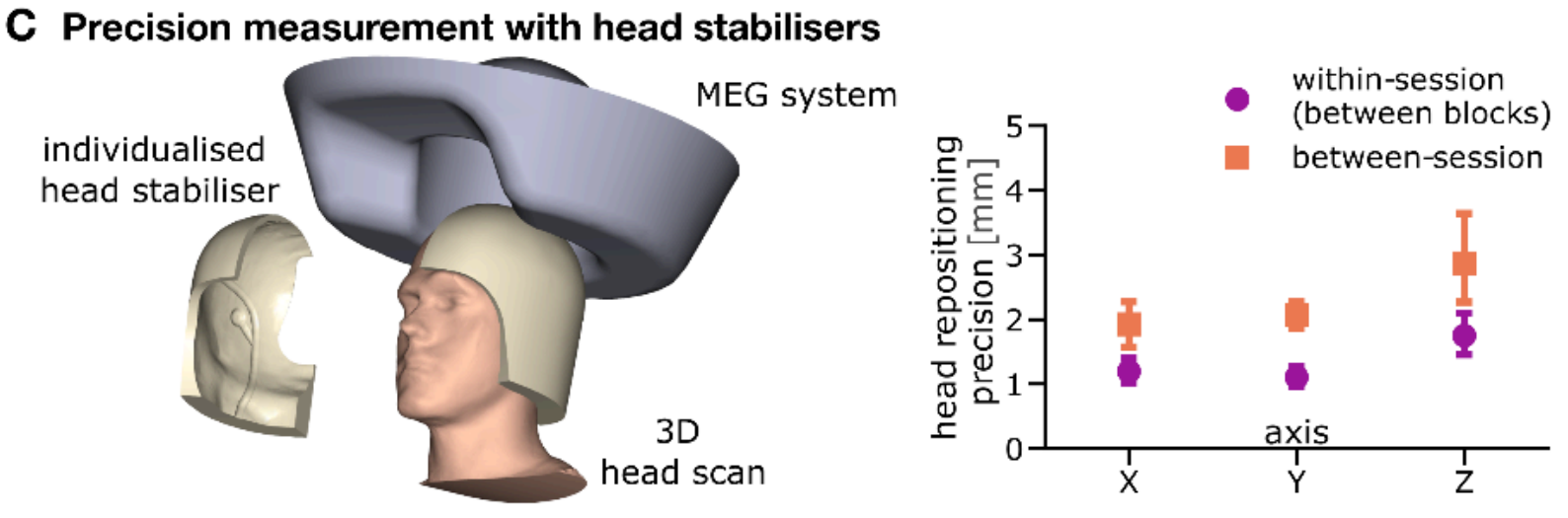

**D** Semantically-balanced natural scene sampling

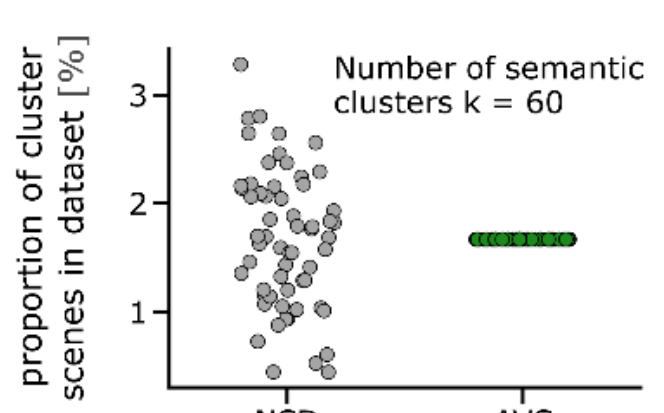

**E** Large-scale diversity of natural scene stimuli

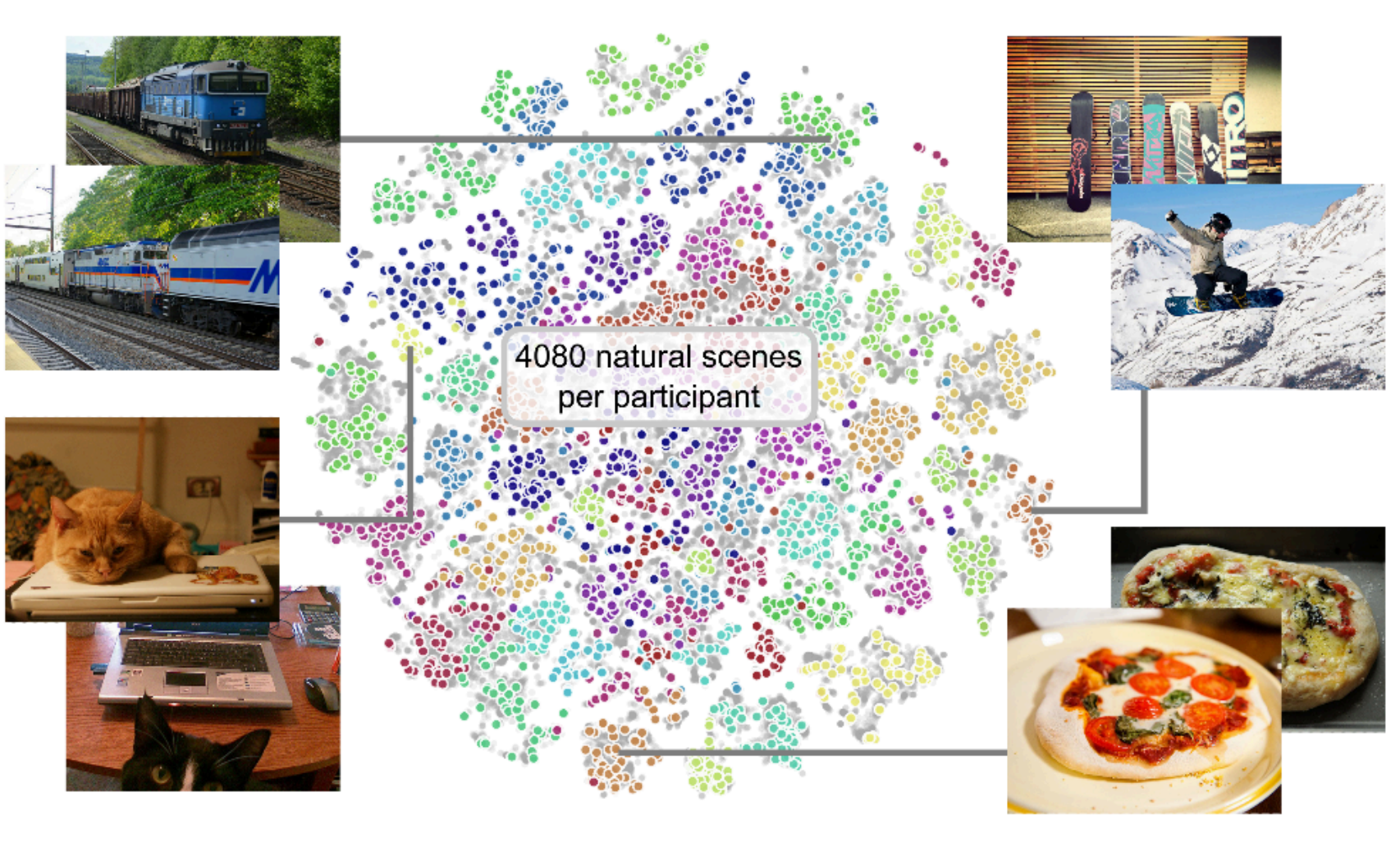

**F** Regression-based eye artifact cleaning

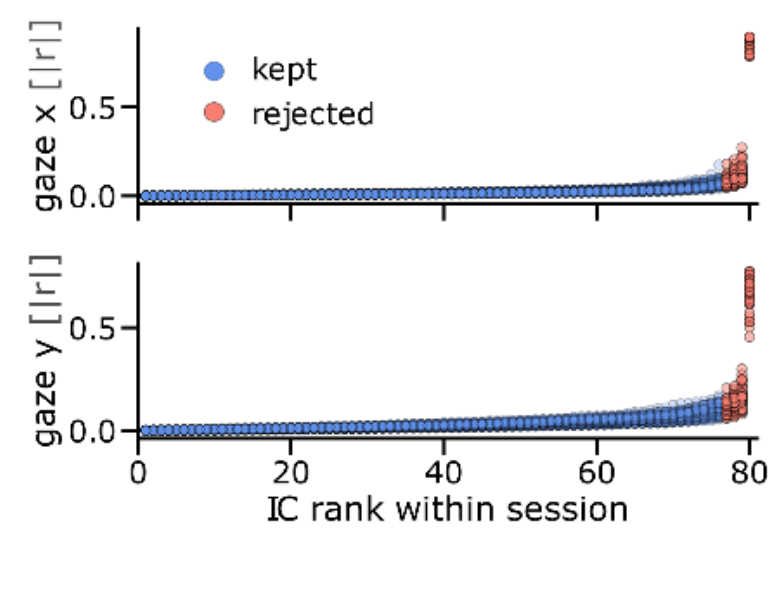

average of rejected ICs

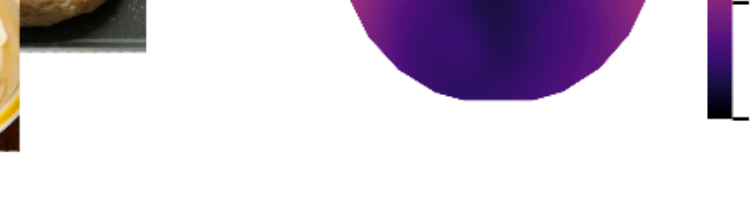

**G** Saccade-aligned neural dynamics at scale

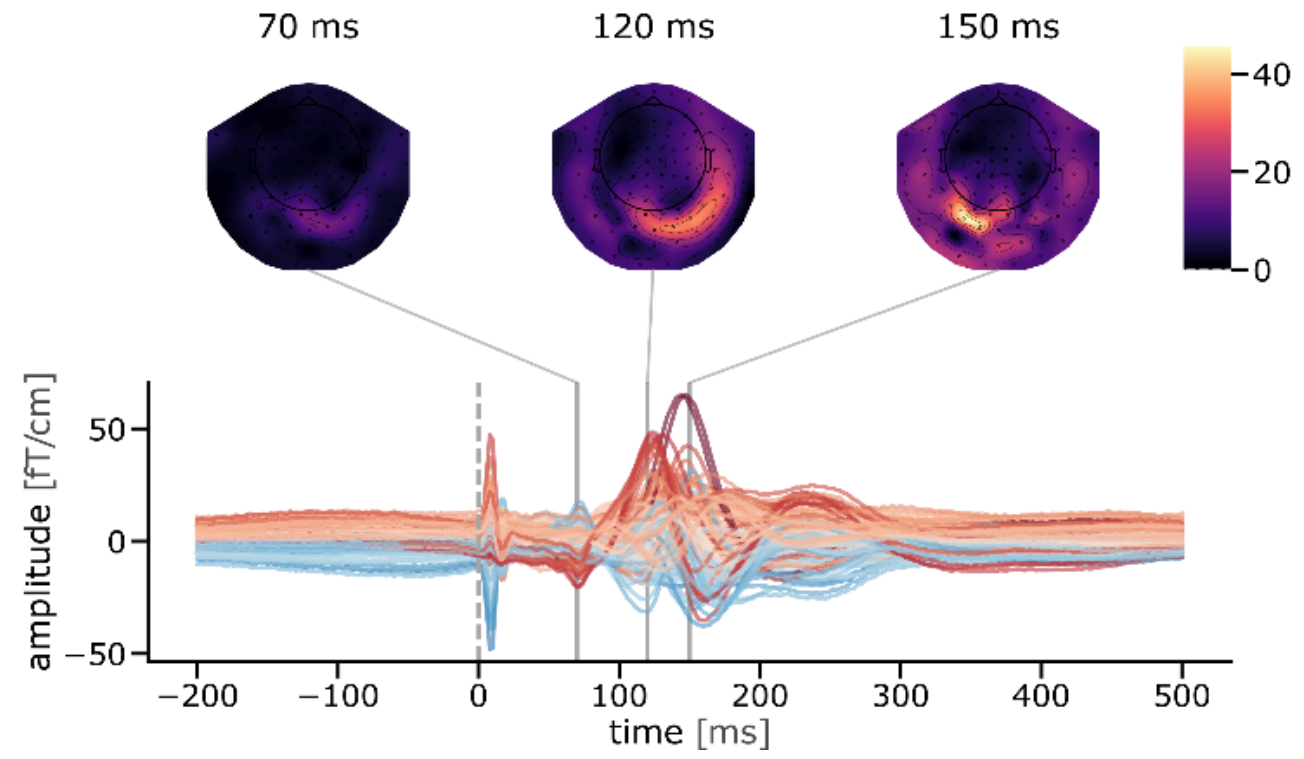

**H** Self-timed saccades structure neural dynamics

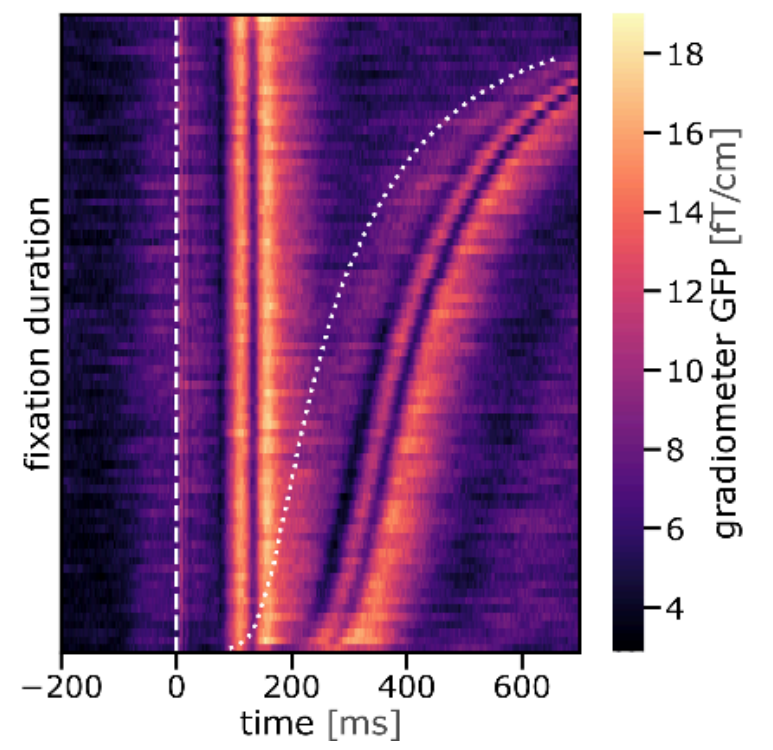

**Figure 1 | The Active Visual Semantics (AVS) dataset: experimental design, data quality, and stimulus diversity. A)** Experimental design. Participants freely viewed 4,080 natural scenes from NSD (Allen et al., 2022) for 4 s each while MEG and eye movements were recorded simultaneously. After 25% of randomly selected trials, participants verbally described the preceding scene. Inset: the three core components of the AVS dataset. **B)** Recording setup (top) and session timeline (bottom) showing the temporal distribution of 10 recording sessions per participant, colour-coded by session number. **C)** Head stabilisation: individualised foam casts were milled from 3D head scans to fit the MEG helmet. Right: within-session (between blocks) and between-session head repositioning precision across the three spatial axes (mean ± 95% CI over subjects, n = 5). **D)** Semantically balanced stimulus sampling. Proportion of scenes per semantic cluster (k = 60) in the full NSD stimulus set (grey) and the AVS subset (green), demonstrating uniform coverage in AVS. **E)** Semantic diversity: t-SNE visualisation of sentence-BERT caption embeddings for all 4,080 AVS scenes, with colours indicating semantic clusters. Example scenes shown at corresponding positions in the embedding space. **F)** Regression-based eye-artefact component identification. Per-session ICA components ranked by absolute correlation of their source time series with horizontal (gaze x) and vertical (gaze y) gaze position; components exceeding the rejection criterion (red) were removed, the remainder retained (blue). On average, 7.8 ± 0.4 components per session were rejected (range 7–8 across 50 sessions; mean $|r|$ = 0.160 ± 0.015 for gaze x, 0.149 ± 0.017 for gaze y). Bottom: gradiometer root-mean-square of the removed artefact, averaged across all rejected components (n = 417, including gaze and cardiovascular), concentrated over frontal and peri-ocular sensors. **G)** Saccade-aligned neural dynamics (Subject 1). Gradiometer responses time-locked to saccade onset (coloured lines: individual channels). Insets: gradiometer GFP topographies at 70, 120, and 150 ms. **H)** Saccade-aligned neural dynamics at scale (Subject 4). Median gradiometer GFP for saccade-locked epochs, binned by subsequent fixation duration (y-axis). The evoked response peaks at ~110 ms; a diagonal ridge (dotted line) tracks the onset of the subsequent saccade and fixation. Dashed white line indicates fixation onset.

# Results

### Massive high-quality sampling of the neural dynamics during active scene viewing

Allowing participants to move their eyes freely relinquishes experimental control over the timing and location of fixations; valid inference therefore rests on tightly constraining the remaining experimental variables (such as stimulus selection and recording quality/parameters) and on sampling sufficient fixation events to estimate neural responses reliably. With this goal, we recorded data from five participants (3 female, 2 male; mean age 27.8 years, SD 2.6) over 10 MEG sessions (median days between sessions: 2; IQR: [1, 5]; range: 1–93 days; Figure 1A, B).

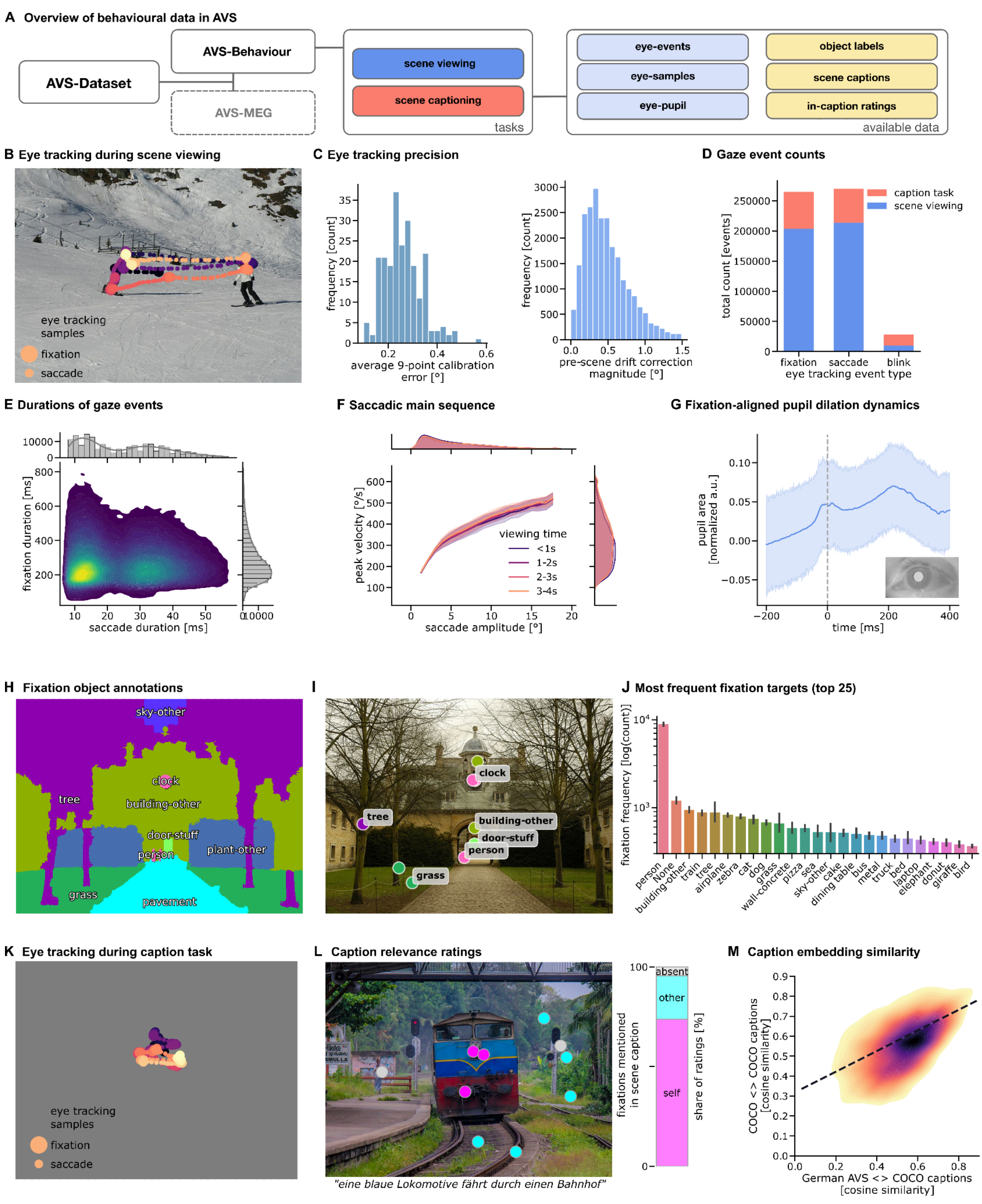

**Figure 2 | Eye tracking data, per-fixation object annotations, and scene captioning behaviour in the AVS dataset. A)** Overview of behavioural data streams available in the AVS dataset. **B)** Example scanpath during scene viewing, with fixations (large points) and saccades (small points) overlaid on a stimulus scene.

**C)** Eye tracking precision: distribution of average 9-point calibration errors (left) and pre-scene drift correction magnitudes (right) across all sessions. **D)** Total gaze event counts (fixations, saccades, blinks) during scene viewing (blue) and the captioning task (red). **E)** Joint distribution of fixation and saccade durations (each fixation paired with its preceding saccade) across all participants, with marginal histograms. **F)** Saccadic main sequence (peak velocity as a function of saccade amplitude), shown separately for four scene viewing time bins within the 4-second exploration period. Shaded areas indicate bootstrapped 95% CI over subjects (n = 5). Marginals show saccade duration and peak velocity distributions. **G)** Fixation-aligned pupil area dynamics (mean ± bootstrapped 95% CI across participants). Inset shows example eye camera image. **H)** MS-COCO and COCO-Stuff (Caesar et al., 2018; Lin et al., 2014) segmentation map for an example scene. **I)** Same scene with fixation locations mapped to their corresponding object category labels. **J)** Fixation frequency (log scale) across the 25 most frequently fixated COCO-Stuff categories. Error bars indicate bootstrapped 95% CI over subjects (n = 5). **K)** Example scanpath during the captioning task for the scene shown in B. **L)** Mentioned-in-caption ratings: example scene with fixation targets classified as referenced in the participant's own caption ("self"), another participant's caption only ("other"), or absent from all captions. Proportions (right) are computed across all participants. **M)** Caption embedding similarity: cross-lingual cosine similarity between German AVS captions and English COCO captions (distiluse-base-multilingual-cased; Reimers & Gurevych, 2019) plotted against within-COCO caption agreement, pooled across all participants. Dashed line shows OLS regression fit.

Sampling natural scenes from internet-derived image collections inherits the upload biases of their sources, for example by over-representing scene types that users share more often. NSD is no exception. To characterise this imbalance, we partitioned sentence-transformer embeddings of its MS-COCO captions (Reimers & Gurevych, 2019) into K-means clusters, setting the cluster count to 60 by maximising the median of the split-averaged per-sample silhouette distribution on held-out scenes across a range of candidate *k* (see Methods). Across the resulting partitioning, per-cluster representation varied substantially (range: 0.44%–3.28% of scenes per cluster; Figure 1D). To prevent this imbalance carrying over into AVS, we sampled stimuli to equalise coverage across the semantic space, drawing 68 scenes uniformly from each of the 60 clusters (4,080 scenes in total; Figure 1D, E). This balanced design reduces the risk that downstream analyses are dominated by the most heavily represented regions of the stimulus space.

## Individualised head stabilisers yield stable between-session repositioning

To enable reliable multi-session source analyses, we stabilised head position with individualised foam casts milled from 3D head scans (Figure 1C). Across sessions, repositioning errors were 1.92 mm (95% CI [1.57, 2.27]), 2.07 mm ([1.86, 2.27]), and 2.87 mm ([2.27, 3.66]) along the X, Y, and Z axes; within-session errors across blocks reached 1.20 mm ([0.88, 1.52]), 1.11 mm ([0.87, 1.42]), 1.75 mm ([1.47, 2.12]), along the X, Y, and Z axes. These errors are well below the 4–5 mm typical of conventional surface- or fiducial-based co-registration (Troebinger et al., 2014) and the deviation thresholds used to trigger re-acquisition in movement-compensation protocols (Stolk et al., 2013). The foam casts thus supported reliable cross-session source reconstruction while preserving the unconstrained viewing the dataset requires.

## Fixation-locked MEG recordings reveal clear evoked responses and systematic temporal structure

A key requirement for the analyses presented in subsequent sections is that active-vision-locked MEG signals contain reliable evoked responses despite the internally-driven timing of gaze events and the potential for oculomotor contamination of the recordings. To address the latter, ocular and cardiac components were removed per session by ICA prior to epoching (Figure 1F; see Methods). Gaze events, such as saccade onset, elicited a robust gradiometer response peaking around 120 ms, with a posterior topography (Figure 1G), confirming that self-generated saccades structure the neural dynamics. Gradiometer global field power (GFP), computed across all saccade event epochs, sorted by subsequent fixation duration (Figure 1H; Subject 4), demonstrates a clear evoked response that peaks at ~110 ms post-fixation onset across the full range of durations. The heatmap further reveals a sharp diagonal ridge corresponding to the evoked response triggered by the subsequent saccade and fixation, with its latency tracking fixation duration precisely.

**Per-scene eye-tracking quality control yields accurate gaze data at scale**

Accurate gaze data are essential for the fixation-locked analyses central to the AVS dataset. Eye-tracking precision was assessed using 9-point calibrations performed at the start of each session and after each break, yielding a mean calibration error of 0.268° (95% CI [0.227, 0.306]; Figure 2C, left). To maintain per-trial accuracy, each pre-scene fixation cross served as a drift correction target. Pre-scene drift corrections had a median magnitude of 0.440° (IQR: 0.270–0.670°; Figure 2C, right), with 92.9% below 1.0° (95% CI [89.4, 96.3]), ensuring stable gaze-recording accuracy throughout data collection.

Across all participants and sessions, the dataset contains 203,356 fixations (mean = 40,671 per subject, 95% CI [37,966, 44,548]), 213,810 saccades (mean = 42,762 per subject, 95% CI [39,684, 46,777]), and 9,954 blinks (mean = 1,991 per subject, 95% CI [1,173, 2,841]) during scene viewing, and an additional 61,624 fixations (mean = 12,325 per subject, 95% CI [7,759, 17,264]), 56,047 saccades (mean = 11,209 per subject, 95% CI [6,403, 16,514]), and 18,097 blinks (mean = 3,619 per subject, 95% CI [2,724, 4,588]) during the caption task (Figure 2D). Together, this yields a large-scale resource of precise eye tracking data that enables in-depth analysis of gaze behaviour during natural scene viewing.

**Gaze event statistics are in line with established properties of natural scene viewing**

Whether participants engage in genuine exploratory behaviour, rather than passive or stereotyped scanning, is not guaranteed by simply allowing eye movements. We therefore computed descriptors of gaze behaviour and compared them to established work from the scene-viewing literature. Fixation durations (n = 199,289; top 2% outliers removed) had a median of 254.4 ms (95% CI [230.8, 273.6]; IQR: 196.0–345.6 ms), and saccade durations (n = 209,541; top 2% outliers removed) a median of 25.2 ms (95% CI [23.4, 28.0]; IQR: 14.2–35.0 ms; Figure 2E), both consistent with established values for free viewing of natural scenes (Devillez et al., 2020; Henderson, 2003; Henderson et al., 2014; Henderson & Pierce, 2008; Nuthmann, 2017; Ramos Gameiro et al., 2017; Wilming et al., 2017).

The saccadic main sequence, the power-law relationship between saccade amplitude and peak velocity, is a stereotyped signature of saccade generation; reductions in saccadic vigour (lower peak velocity for a given amplitude) are a well-established marker of fatigue and declining arousal (Bahill et al., 1975; Gibaldi & Sabatini, 2021; Harris & Wolpert, 2006). In our task, the main sequence remained stable across viewing time (Figure 2F). Log-log slopes did not differ significantly between time bins (mixed-effects model with random intercepts per subject; all $p > .29$), with mean slopes ranging from 0.409 to 0.417 (with a mean correlation of $r = 0.87$, 95% CIs [0.81, 0.92], n = 5 subjects). This stability indicates that oculomotor dynamics were preserved across the full 4 s viewing period, consistent with sustained active exploration rather than progressive fatigue or reduced engagement.

A further window into visual cognition has been reported in the form of pupil dilation responses (Grujic et al., 2024; Joshi & Gold, 2020; Koevoet et al., 2025). In our data, the fixation-locked pupil area dynamics showed a transient dilation shortly before fixation onset, followed by gradual dilation in response to the subsequent fixation target (Figure 2G). This suggests visually selective, temporally resolved pupil response data that allows for the analysis of pupil dynamics during natural scene viewing at scale.

**Per-fixation object labels capture the distribution of fixated content during natural scene exploration**

To further enhance the usefulness of our dataset, we linked each fixation to the visual and semantic content at the respective location. By mapping gaze positions onto MS-COCO and COCO-Stuff instance segmentations (Caesar et al., 2018; Lin et al., 2014), each fixation was assigned the category label of the underlying object segmentation (Figure 2H, I). Of 203,356 total fixations, 97.0% received an object label across 171 unique categories, with "person" targets dominating. While this is consistent with the well-established priority of person and face content in natural scene viewing (Birmingham et al., 2008; Yarbus, 1967), the present fixation counts reflect observed fixation frequency rather than fixation density relative to category availability over scenes. The remaining most frequently fixated categories span a variety of buildings, animals, vehicles, and food items. This large number of per-fixation object labels enables category-level analyses of neural data recorded during active viewing, for instance, the fixation-based representational similarity analyses (see Figure 3).

**Scene captions capture semantic content and link fixation behaviour to scene understanding**

To encourage active viewing behaviour that serves semantic scene understanding, after 25% of trials, participants were tasked with verbally describing the preceding scene (Figure 2K, L). To assess caption quality, we compared participant-generated captions (in German) to the original MS-COCO (Lin et al., 2014) annotations (in English) using multilingual sentence embeddings (distiluse-base-multilingual-cased; Reimers & Gurevych, 2019). Mean cross-lingual cosine similarity was 0.53 (95% CI [0.51, 0.55]), compared to a within-COCO self-similarity of 0.59 (95% CI [0.59, 0.59]). Cross-lingual similarity scaled with within-COCO caption agreement ($b$ = 0.61, 95% CI [0.58, 0.64], $p < .001$; Figure 2M), confirming that the German AVS captions reliably trace the semantic content captured by the original English annotations despite the language difference.

To explicitly link the gaze behaviour to the downstream captioning task, two independent raters classified each fixation target (n = 45,193 per rater; 5 subjects, 1,020 scenes) as referenced in the participant's own caption ("self"), in another participant's caption only ("other"), or absent from all captions (Figure 2L). On average across raters, 82.81% ± 12.53% (mean ± SD) of fixation targets were classified as "self", 13.06% ± 12.91% as "other", and 4.13% ± 0.38% as absent. The high proportion of self-referenced fixation targets provides a behavioural link between gaze allocation and subsequent memory-based report.

**A** **Per-fixation ANN to MEG encoding**

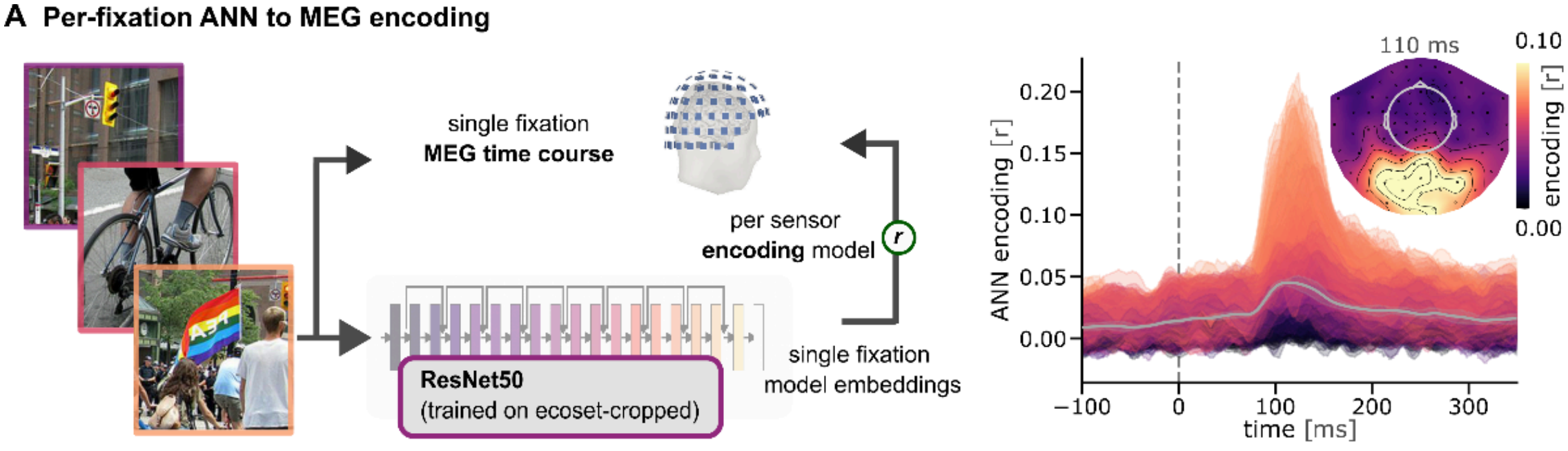

**B** **Active vision dynamic representaion similarity analysis (dRSA)**

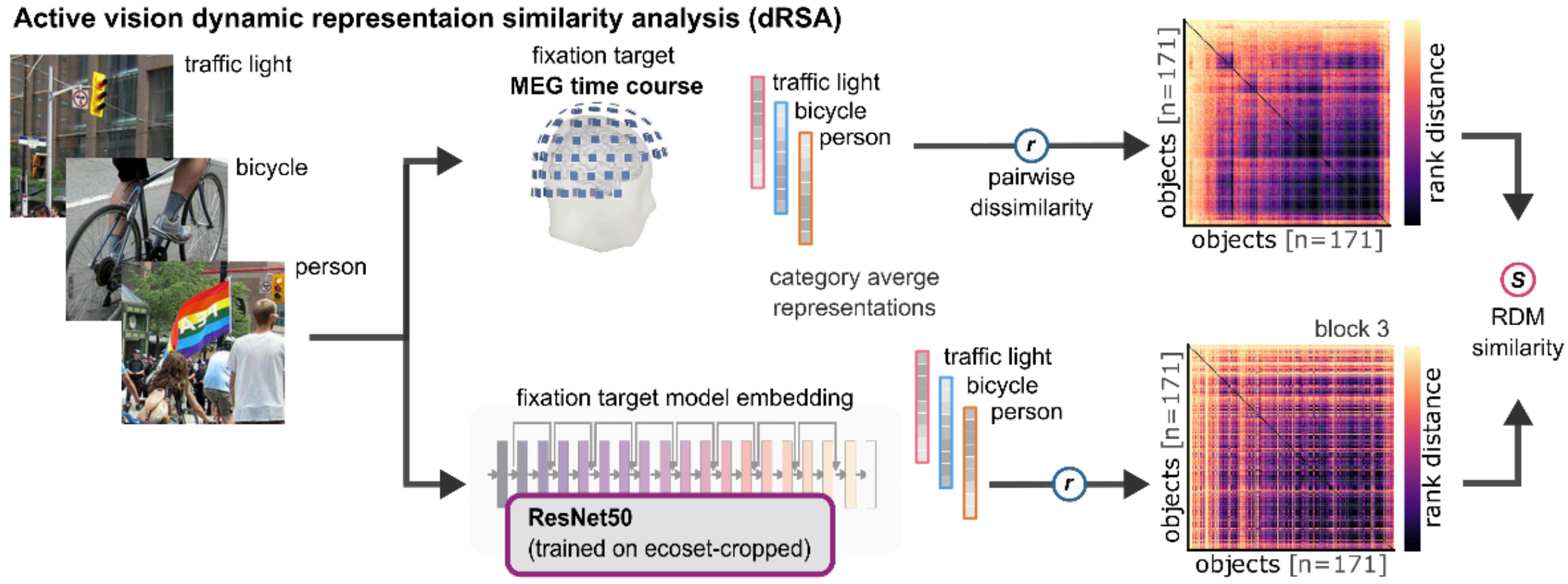

**C** **Inter-subject similarity of fixation-based RDMs**

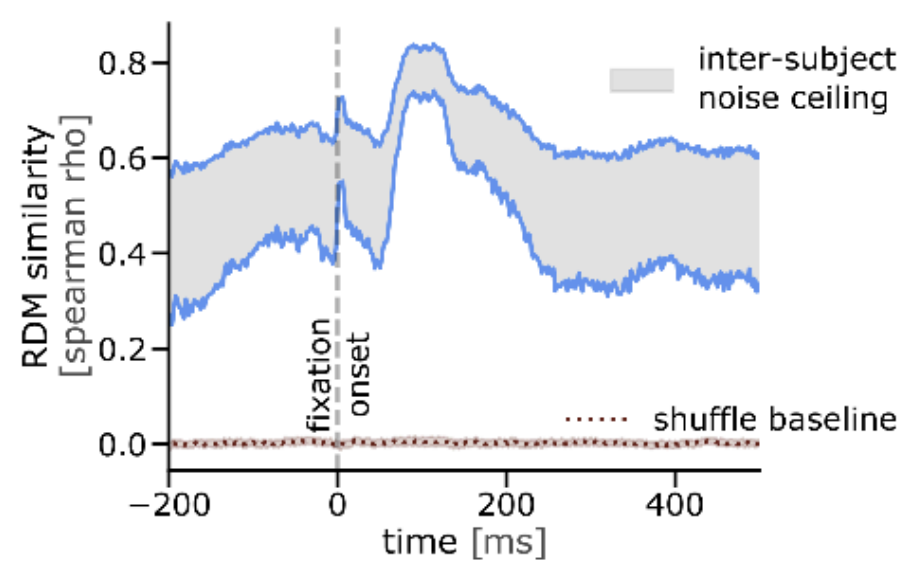

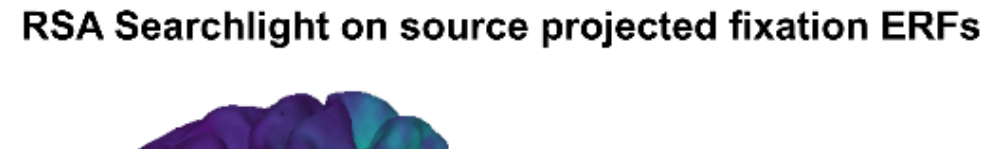

**D** **Layer-wise fixation based dRSA**

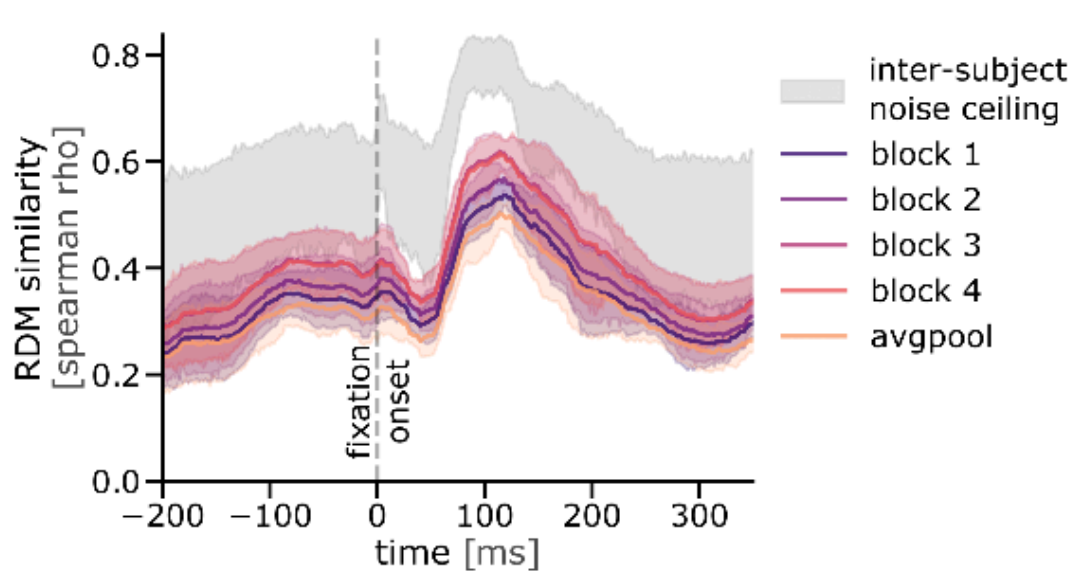

**E** **RSA Searchlight on source projected fixation ERFs**

**F** **Noise ceiling of RSA Searchlight**

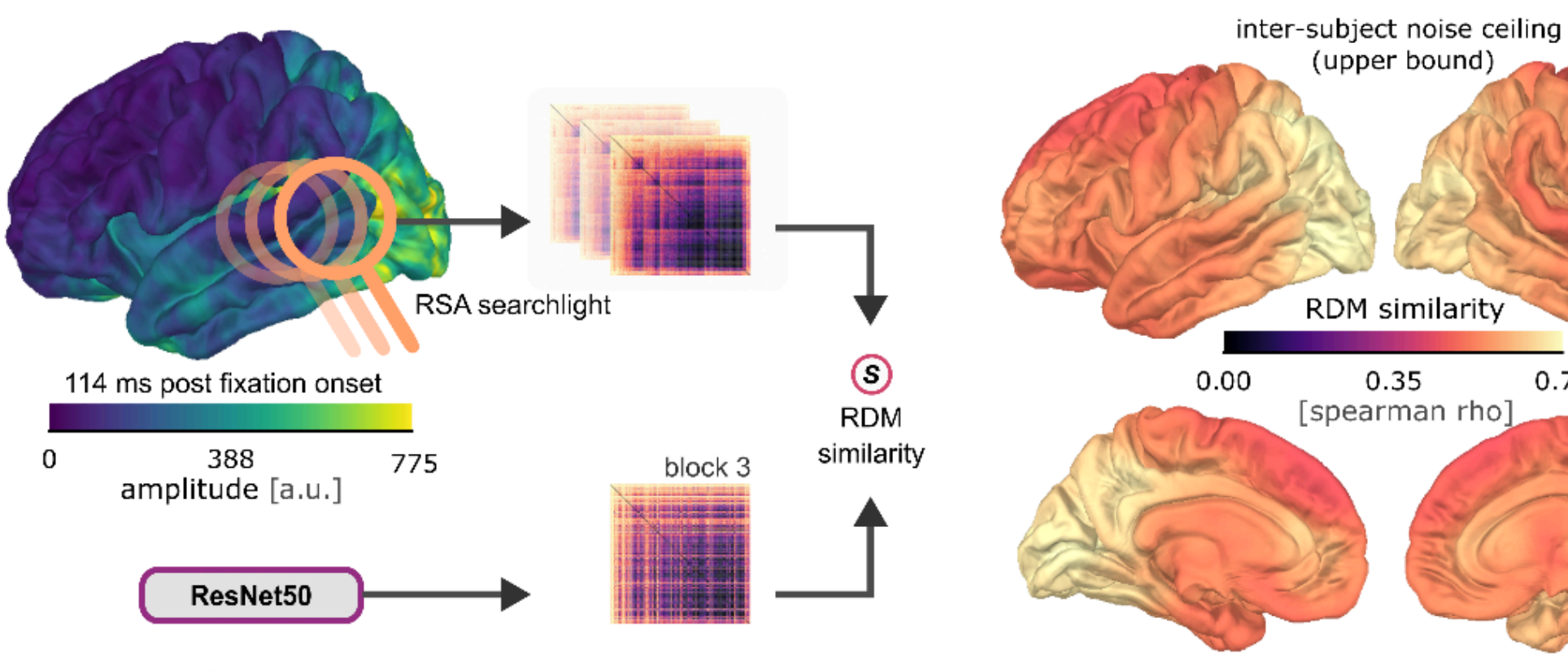

**G** **ANN vs. MEG source searchlight**

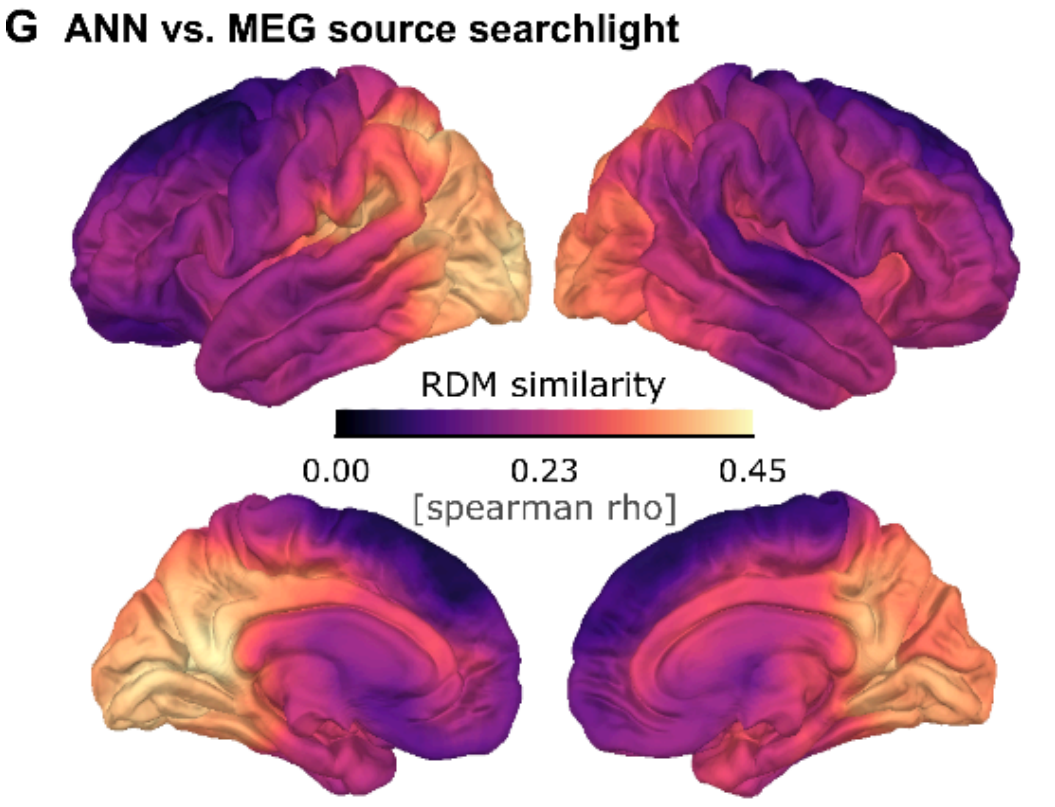

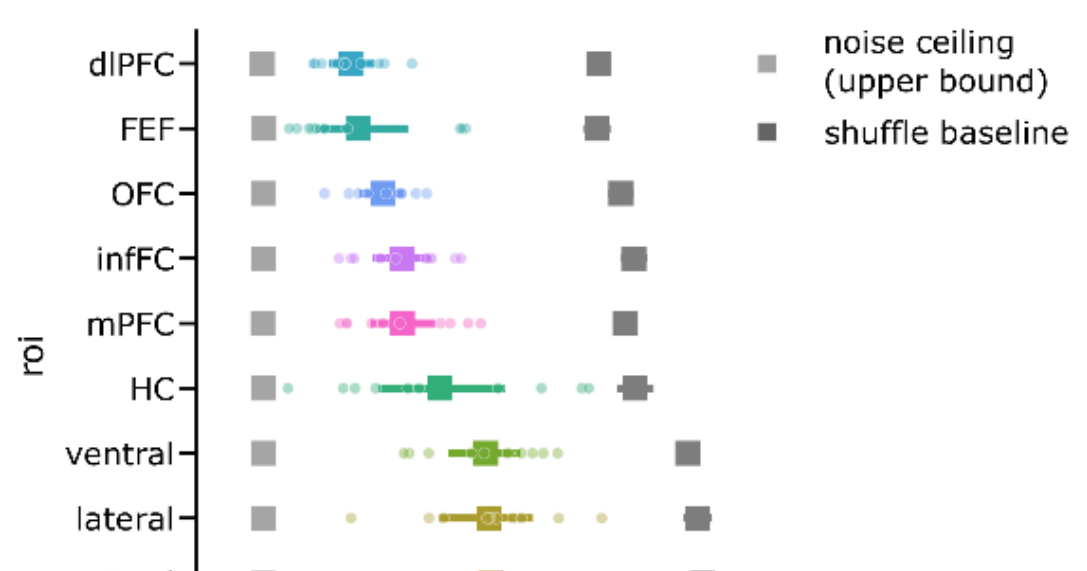

**Figure 3 | MEG sensor and source signals encode fixated visual content during active scene viewing. A)** Per-fixation encoding. For each fixation, 112 × 112 pixel image patches centred on the gaze position were embedded via ResNet50 (trained on crop-augmented ecoset). Per-channel ridge regression models predicted fixation-locked gradiometer responses from these embeddings (80/20 train-test split, scene-grouped). Right: time-resolved encoding accuracy (Pearson's r; coloured lines: individual channels; black: mean across channels; shading: bootstrapped 95% CI across subjects). Inset: sensor topography at peak latency (110 ms). **B)** Fixation-aligned dynamic RSA (dRSA). Gradiometer responses were averaged across all fixation events within each of 171 object categories (minimum 10 fixations per category). Pairwise correlation distances yielded time-resolved neural RDMs; a static model RDM was computed from ResNet50 block 3 embeddings. **C)** Time-resolved neural–model RDM similarity (Spearman's ρ) for ResNet50 block 3, with inter-subject noise ceiling (grey shading; blue lines: lower and upper bounds) and category-shuffle baseline (dotted). **D)** Layer-wise dRSA for ResNet50 blocks 1–4 and avgpool (shading: bootstrapped 95% CI across subjects; grey: inter-subject noise ceiling). **E)** Source-space RSA searchlight approach. Fixation-locked category ERFs were source-reconstructed using dSPM and morphed to fsaverage. Left: cortical source amplitude at 114 ms post-fixation onset. A searchlight (20 mm radius, geodesic distance along the cortical surface) computed local neural RDMs, which were correlated with the block 3 model RDM (Spearman's ρ). **F)** Inter-subject upper-bound noise ceiling of the RSA searchlight on the cortical surface (lateral and ventral views). **G)** Cortical surface map of neural–model RDM similarity (left) and ROI-level summary (right; point plots with error bars across 5 subjects × 2 hemispheres, individual data points shown; grey: upper-bound inter-subject noise ceiling).

## Individual fixations carry fixation target-specific signal during active vision

Having validated the eye tracking recording quality, we next test neural signal quality and start with the probably most demanding case. We investigate whether the visual content of a single, self-generated fixation is recoverable from the MEG signal. Unlike static paradigms, active vision offers neither randomised onsets nor stimulus repetitions, the two features conventional analyses use to raise signal-to-noise. Each fixation is a unique onset embedded in a continuous sequence of exploration, seen once and never averaged.

We fitted per-channel encoding models predicting fixation-locked gradiometer responses from ResNet50 embeddings (average pooling layer; He et al., 2016; trained on a cropped version of ecoset; Mehrer et al., 2021) of image patches centred on each fixation (Figure 3A). Encoding correlation rose sharply after fixation onset and peaked at 114 ms (mean r = 0.045, bootstrapped 95% CI [0.042, 0.049]; across subject-level channel averages, N = 5), strongest at posterior sensors that reach encoding correlations of about r ~ 0.2 (Figure 3A, inset).

This latency and topography place a large part of the effect within the early visual response. The 114 ms peak coincides with the early, posterior encoding responses reported for stimulus onset-locked controlled paradigms using comparable DNN-feature encoding models: 75 to 135 ms over early visual cortex in MEG (Seeliger et al., 2018) and a 110 ms peak over occipito-parietal channels in EEG (Gifford et al., 2022). Both report those accuracies on encoding of trial-averaged responses, and Seeliger et al. show that averaging repetitions raises encoding performance drastically (up to a factor of two). Without any form of averaging, or explicitly modelling the sequential nature of the active vision MEG data, the reported encoding performance can be seen as a lower bound: what survives the hardest test the dataset allows, single-channel, single-trial model, without the noise reduction conventional designs afford. Its presence establishes that active-viewing MEG retains the fixation-specific visual information downstream analyses depend on.

### Object category structure emerges in fixation-locked neural representations

The single fixation MEG data in AVS carries encodable visual features. Yet, the single-fixation encoding result rests on a heavily parametrised mapping: one linear model per channel, fitted from DNN features to responses. Encoding models were therefore met with the concern that the fit, not the model, does most of the work of matching the brain (Doerig et al., 2023; Storrs et al., 2021). Representational similarity analysis (RSA) offers a complementary approach that, by default, fits nothing. Its power rests in comparing neural and model representations through their pairwise dissimilarity structure alone (Kriegeskorte et al., 2008). However these pairwise comparisons usually are derived from category average neural response patterns. Yet here in the active vision case, such an average would operate over fixations sharing only a category label, not identical input. Whether category-level averaging recovers reproducible structure, rather than washing out under the heterogeneity of natural scenes and gaze targets, is therefore an empirical question. To test this, we grouped fixation-locked responses by object category ($n = 171$; Caesar et al., 2018) and asked how much category-level structure is reproducible across participants (inter-subject noise ceiling) and how much aligns with an image-computable deep neural network model (ResNet50 embeddings computed from the fixation targets, Figure 3B).

We start by quantifying the reliability of the category-level neural RDMs. For this, we estimated inter-subject noise ceilings by correlating each participant's neural RDM with the group-average RDM (upper bound) and the leave-one-out group average (lower bound; Figure 3C). The upper bound of the noise ceiling peaked at Spearman's $\rho = 0.838$ (114 ms post-fixation onset), indicating that category-level representational structure is highly consistent across participants and establishing a high signal reliability against which to benchmark models. RDM similarity with the ANN rose shortly after fixation onset and was highest for block 3 of ResNet50, peaking at $\rho = 0.620$ (mean across 5 subjects; bootstrapped 95% CI [0.581, 0.651]) also at 114 ms. The coincident timing of the ceiling and model peaks (both 114 ms) further indicates that the model features capture category information in the MEG signal at the point where it is most reliably expressed across participants. Notably, the RDM alignment was already elevated at fixation onset (e.g., block 3: $\rho = 0.408$, 95% CI [0.336, 0.465]), likely reflecting temporal overlap between adjacent fixation epochs, successive fixations on same-category objects, or the spatial co-occurrence of visually or semantically similar objects within natural scenes. A linear mixed-effects model with random intercepts per subject confirmed that block 3 explained more variance than block 1 at peak time ($\beta = 0.088$, 95% CI [0.076, 0.100], $p < 0.001$; $N = 5$). Together, these results show that category-level averaging survives the heterogeneity of natural viewing, recovering representational structure that is highly reproducible across participants and closely tracked by (mid-level) ANN features, and thus provides a reliable basis for benchmarking image-computable models of single-fixation MEG.

### Source-space RSA localises fixation-based category representations to visual cortex

Having established high inter-subject reliability and overall data quality at the sensor level, we turned to source space to (i) test whether this reliability localises to regions expected for visual processing, (ii) characterise that localisation under naturalistic active viewing, and (iii) establish that the dataset supports source-space analyses of active vision more generally. Using each participant's structural MRI, we source-reconstructed fixation-locked category ERFs with dSPM and individual BEM forward models, applied session-specific noise covariance, and morphed source estimates to fsaverage. A geodesic searchlight (20 mm radius) was applied at the group-level RSA peak latency (114 ms post-fixation onset; Figure 3E), and categories with fewer than 10 fixations were excluded.

Mirroring the sensor-level RSA, for each searchlight we computed the inter-subject upper-bound noise ceiling across categories seen by all five participants. This quantifies how much fixation-locked category structure is shared across participants independent of any model. The average per-vertex searchlight noise ceiling for a set or ROIs followed a posterior-to-anterior gradient (Figure 3F): it was highest in early visual ($NC_{upper}$ = 0.66), parietal (0.65), lateral (0.65) and ventral cortex (0.63), intermediate in hippocampus (0.55), infFC (0.55), mPFC (0.54) and OFC (0.53), and lowest in FEF (0.50) and dlPFC (0.50). Consistent category information was thus present throughout but most reliable in occipito-temporal and ventral visual cortex.

To ask how much of this reliable structure a standard feedforward model captures, we correlated each vertex's searchlight RDM with an ANN model RDM (ResNet50 block 3) and averaged within ROI and hemisphere per participant. Model-brain alignment followed the noise ceiling reliability gradient. It was strongest in early visual, parietal, lateral and ventral cortex (Spearman's $\rho$ = 0.33 to 0.37) and weakest in frontal ROIs ($\rho$ = 0.13 to 0.14; Figure 3G). All ROIs exceeded a per-searchlight label-shuffled baseline (one-sided one-sample t-tests across participants, BH-FDR corrected across ROIs; all $p_{FDR} < 0.02$). Notably, the model captured a larger share of the available ceiling in posterior regions (just over half in visual cortex) than in frontal regions (roughly a quarter), leaving substantial reliable variance unexplained even where alignment was best.

Together, these analyses establish that fixation-locked category structure is reliable across participants and spatially organised along a posterior-to-anterior gradient that peaks in visual cortex. A standard feedforward model recovers part but not all of this structure, following the same gradient.

## Discussion

The Active Visual Semantics (AVS) dataset shows that the repertoire of analyses established in computational visual neuroscience, from single-epoch encoding to time-resolved representational similarity analysis with noise-ceiling estimation in sensor and source space, can be successfully applied to MEG data collected during active, self-timed natural viewing. This rests on three features of the dataset: its scale (more than 200,000 fixation epochs across 10 sessions per participant), the signal quality afforded by individualised head stabilisation and high-acuity eye-tracking, and the semantic and visual diversity of the 4,080 naturalistic scenes.

AVS complements existing large-scale neuroimaging resources such as NSD (Allen et al., 2022) and THINGS-data (Hebart et al., 2023), which have proven transformative for understanding scene and object representations under controlled viewing. AVS extends this programme into active vision, where the observer's own behaviour determines what is seen, when, and for how long. The addition of per-fixation object labels, a semantic captioning task, and the millisecond-resolved temporal structure of MEG demonstrates that AVS is in a unique position to address questions about the neural interface between perception, action, and higher-level semantic processing, giving empirical traction to the notion of studying visual intelligence in action.

Several constraints warrant consideration. (i) The intensive within-subject design prioritises statistical power for per-fixation analyses over population-level inference, following the rationale of other densely sampled datasets (Allen et al., 2022; Hebart et al., 2023; Naselaris et al., 2021);

generalising specific effect sizes would require additional participants. (ii) The captioning task on 25% of trials may have biased eye movements toward features relevant for verbal description. However, free viewing is itself reported to be best described by a scene-comprehension objective: unconstrained fixations resemble those made when describing scenes, and differ from those made when searching or counting (Murlidaran & Eckstein, 2025). A captioning task could therefore be argued to be the default mode of natural viewing, while reducing the uncertainty about the participant's current set of objectives followed while viewing freely without explicit instruction. (iii) More fundamentally, head-fixed active vision, while substantially more naturalistic than passive viewing, remains a compromise between signal quality and ecological validity. With advances in mobile and wearable neuroimaging, it may become possible to carry high-resolution neural signals into yet more unconstrained paradigms, but it should be noted that one of the strengths of AVS is the large semantic coverage, which is unlikely to be possible in real-world measurements.

The dataset has already yielded findings that illustrate its broader potential. For example, Sulewski et al. (2026) showed that fixation durations are better predicted by memory encoding than by visual processing demands, with longer fixations co-occurring with enhanced theta-gamma coupling in frontal and hippocampal regions. Amme, Sulewski et al. (2026) demonstrated that saccade-related events, not fixation onsets, best explain the latency and topography of early visual cortical responses during naturalistic viewing. Ongoing work examines how neural oscillations encode context-based informativeness during free viewing (Bai et al., n.d., in prep) and whether cross-saccadic predictions of neural responses rely on high-level rather than low-level visual features (Kämmerer, Amme et al., in prep.). Across these studies, a common theme is emerging: when observers control their own visual sampling, internally generated signals reshape neural dynamics in ways that passive paradigms neither capture nor predict. This is consistent with a broader move in the field towards studying cognition under naturalistic conditions (Borovska & de Haas, 2024; Carvalho & Lampinen, 2026; Cisek & Green, 2024; Huber-Huber & Lange, 2025; Popov, 2026), motivated by growing awareness that the assumptions inherited from traditional paradigms may not generalise to natural behaviour. Understanding vision as an active process, where the observer determines what is sampled and when, is a necessary complement to controlled paradigms, and this dataset is built to enable it.

## Methods

### Participants

Five healthy, right-handed participants (3 female, 2 male; mean age = 27.8 years, SD = 2.6), all native German speakers with normal or corrected-to-normal vision, took part in the study. All participants provided written informed consent. The study was approved by the ethics committee of the University of Leipzig and conducted in accordance with the Declaration of Helsinki.

### Stimuli and experimental design

Stimuli comprised 4,080 natural scenes from the Natural Scenes Dataset (NSD; Allen et al., 2022), including the shared-1000 subset presented to all NSD participants and the special-100 subset that maximally spans the NSD semantic space. To ensure broad semantic coverage, we computed sentence-BERT embeddings ("paraphrase-multilingual-mpnet-base-v2"; Reimers & Gurevych, 2019) of the five MS-COCO captions available per scene (Lin et al., 2014), averaged the five embeddings, partitioned the resulting space into 60 clusters via K-means. The number of clusters was selected using a cross-validated silhouette criterion. For each candidate k, we drew repeated random 90/10 train/test splits, fitted K-means on the training split, assigned held-out test scenes to their nearest centroid, and computed per-sample silhouette values on the test set. We averaged these per-sample distributions across splits and took the median as the selection criterion, choosing the k that maximised it. We then sampled 68 scenes uniformly from each of the 60 clusters. All participants viewed the same stimulus set in individually randomised order.

Scenes were cropped and resized to 947 × 710 pixels, preserving aspect ratio, and displayed on a 41.6 × 31.2 cm screen (1024 × 768 pixels) at a viewing distance of 70 cm, subtending 28.5° × 21.6° of visual angle. Each scene was shown for 4 s, followed by a 1–2 s inter-stimulus interval. On 25% of trials, selected pseudorandomly but fixed across participants, a dark grey microphone appeared on a grey background (50%) for 1 second, which indicated the onset of the verbal scene description task. This was followed by an empty grey screen presented during the 8 s response window.

### MEG acquisition

Brain activity was recorded using a 306-channel whole-head MEG system (Elekta Neuromag TRIUX, Elekta Oy, Helsinki, Finland) with 102 magnetometers and 204 planar gradiometers, sampled at 1000 Hz with an online bandpass of 0.1–330 Hz. Five head-position indicator (HPI) coils tracked head position continuously. Before each session, the participant's head shape was digitised using a Polhemus FASTRAK system. Head stabilisation was achieved using individually fitted foam casts, milled from 3D head scans to fill the space between the participant's head and the MEG helmet, allowing natural viewing behaviour while minimising head movement.

### Eye tracking

Eye movements were recorded with an EyeLink 1000 system (SR Research Ltd., Ottawa, Canada) at 1000 Hz, calibrated using a 9-point grid. A fixation cross preceding each scene served as drift correction; the next scene appeared only after successful correction. Gaze position data were aligned with MEG recordings via triggers sent at trial onset. Saccades and fixations were detected using a velocity-based algorithm (Engbert & Mergenthaler, 2006): for each session, samples exceeding five standard deviations of the sample-wise velocity were classified as saccades; all remaining non-blink samples were classified as fixations. Fixations shorter than 50 ms or longer than 1000 ms and the last fixation per trial were excluded from analysis.

### MEG preprocessing

MEG data were processed using MNE-Python (Gramfort et al., 2013). Temporal signal space separation (tSSS) with movement compensation was applied to suppress external interference (MaxFilter, Elekta Oy). Data were bandpass filtered between 0.2 and 200 Hz and resampled to 500 Hz. Independent component analysis (ICA; FastICA) was applied per session to 80 components extracted from a 1–40 Hz bandpass-filtered copy of the concatenated MEG data. Eye-movement artefact components were identified by computing the Pearson correlation between each IC source time series and the continuous horizontal (gx) and vertical (gy) gaze position signals, epoched around scene onsets. Components falling within the top 5% of absolute correlation with either gaze axis were excluded (union criterion), removing on average 7.8 ± 0.4 ocular components per session (mean $|r| = 0.16$ for gx, 0.15 for gy; range 7–8 per session across 50 sessions). ICA was applied to the broadband data prior to epoching. The MEG data was epoched around fixation onsets (−500 to

+800 ms) based on eye-tracking event markers aligned at each scene onset. No baseline correction was applied.

### Head position analysis

Head-repositioning precision was quantified from continuous HPI coil recordings. For each run, the median position was computed from the initial 10 samples. Within-session repositioning error was defined as the mean-corrected standard deviation of these median positions across runs within a session. Between-session repositioning error was computed as the mean-corrected standard deviation across session-initial positions.

### Fixation target object labeling

Each fixation was assigned an object category label by mapping gaze positions in image coordinates onto MS-COCO and COCO-Stuff instance segmentations (Caesar et al., 2018; Lin et al., 2014) yielding 171 unique category labels spanning both discrete objects and scene regions. Where multiple segmentation masks overlapped at a fixation location, the category with the smallest mask area was assigned. Fixations falling outside all segmented regions were labelled as "None".

### Caption similarity analysis

Participant-generated scene descriptions (in German) were compared to the original MS-COCO annotations (in English) using multilingual sentence embeddings (distiluse-base-multilingual-cased; Reimers & Gurevych, 2019). For each scene, cosine similarity was computed between the participant's caption embedding and the mean embedding of the five MS-COCO captions. Within-COCO self-similarity was computed analogously by comparing each MS-COCO caption to the mean of the remaining four. The relationship between cross-lingual similarity and within-COCO caption agreement was assessed using OLS regression with standard errors per subject.

### Fixation-aligned encoding analysis

For each fixation, a 112 × 112 pixel image patch (~3,5° visual angle) centred on the gaze position was extracted from the stimulus scene. Patches were embedded using the global average pooling layer of a ResNet50 trained on crop-augmented ecoset images (565 categories; Mehrer et al., 2021). Per-channel ridge regression models were trained to predict gradiometer responses at each timepoint from these embeddings. Regularisation was selected via cross-validated ridge regression (RidgeCV; 10 α values log-spaced from $10^{-3}$ to $10^{10}$, 3-fold internal CV). Training and test sets were defined by a single 80/20 split with scene-based grouping, ensuring that fixations from the same scene did not appear in both sets. Encoding accuracy was assessed as the Pearson correlation between predicted and observed sensor responses.

### Representational similarity analysis

**Sensor-space dynamic RSA.** Fixation-locked gradiometer responses were averaged across all fixation events within each of 171 object categories. Category-level representational dissimilarity matrices (RDMs) were computed at each timepoint using correlation distance. Model RDMs were constructed from pairwise correlation distances between mean ResNet50 activations (layers 1–4 and avgpool) for each category. Neural-to-model RDM correspondence was assessed using Spearman correlation. Inter-subject noise ceilings were estimated by correlating each participant's neural RDM with the group-average RDM (upper bound) and with the leave-one-out group average (lower bound). A category-shuffle baseline was computed by randomly permuting category labels to confirm that correspondence was driven by category-specific content.

**Source-space RSA searchlight.** To localise category-specific representational structure on the cortical surface, we source-reconstructed fixation-locked category ERFs using dSPM implemented in MNE-Python (Gramfort et al., 2013). Individual forward models were computed using three-layer BEMs based on FreeSurfer cortical reconstructions (Fischl, 2012), with source spaces at 4098 vertices per hemisphere (ico4). Session-specific noise covariance was estimated from 400 ms pre-scene baseline periods pooled across all 10 sessions using Oracle Approximating Shrinkage (OAS). The inverse solution used loose = 0.2, depth = 0.8, and an assumed SNR of 3 ($\lambda^2 = 1/9$). Individual source estimates were morphed to fsaverage (ico5, smooth = 5).
Category ERFs were averaged over a ±10 ms window centred on the group-level sensor-space dRSA peak (114 ms post-fixation onset), yielding a single activation cortical surface map per category. Categories with fewer than 10 fixations were excluded. A geodesic searchlight (20 mm radius) was applied across the cortical surface using MNE-RSA (Van Fliet et al., 2025). Local neural RDMs were computed using correlation distance and compared to the ResNet50 block 3 model RDM via Spearman correlation. The inter-subject noise ceiling was computed by correlating each participant's neural RDM with the group-average RDM, restricted to categories observed in all five participants. For ROI-level summaries, vertex-wise searchlight values were averaged within each ROI, yielding 10 values per ROI (5 subjects × 2 hemispheres). Visual ROIs (early, lateral, ventral, parietal) were defined following the NSD cortical ROI scheme (Allen et al., 2022). Frontal regions (dlPFC, FEF, OFC, mPFC, infFC) and hippocampus were defined as compound regions from the Glasser parcellation (Glasser et al., 2016).

To establish a chance-level reference for the searchlight RSA, we constructed a category-label permutation null: for each participant, the geodesic searchlight (20 mm radius) was recomputed with the category labels of the model RDM permuted (100 permutations), and because permuting the rows and columns of the square model RDM by the same permutation is equivalent to permuting the underlying category labels, each vertex's local neural RDM (correlation distance) was computed only once and correlated (Spearman) with the observed model RDM and with each of the 100 permuted model RDMs, yielding an observed searchlight map and a per-vertex null distribution; for ROI-level inference, the group-mean null RSA was averaged within each ROI (separately per hemisphere) to define a per-ROI shuffle baseline, displayed with a bootstrapped 95% CI over hemispheres analogous to the noise ceiling, and category-specific representational structure was tested per ROI with a one-sample t-test across participants ($n = 5$) on ROI-mean RSA (averaged across hemispheres) versus the shuffle baseline (one-sided, RSA > baseline), Benjamini–Hochberg FDR-corrected across ROIs.

### Data and code availability

The AVS dataset is publicly available under a CC BY 4.0 license via the AWS Open Data Program supported by an AWS Open Data Sponsorship Program grant and documented at https://kietzmannlab.uni-osnabrueck.de/avs/. The dataset includes raw and preprocessed MEG data, eye-tracking data with fixation and saccade annotations, per-fixation object category labels, transcribed participant scene captions, stimulus presentation logs with timing information, and individual anatomical MRIs (defaced). Analysis code for all results presented in this paper is available at https://github.com/KietzmannLab/pyavs.

**Competing interests**

The authors declare no competing interests.

**Author Contributions**

Conceptualization: P.S., P.K., T.C.K.; Data curation: P.S., C.A.; Formal analysis: P.S.; Funding acquisition: T.C.K., M.N.H., P.K.; Investigation: P.S., P.K., T.C.K.; Methodology: P.S., P.K., T.C.K., M.N.H.; Project administration: P.S., P.K., T.C.K., M.N.H.; Resources: M.N.H., T.C.K.; Software: P.S., C.A.; Supervision: P.K., T.C.K.; Validation: P.S., C.A.; Visualization: P.S.; Writing – original draft: P.S., P.K., T.C.K., M.N.H.; Writing – review & editing: All authors.

**Acknowledgements**

We thank Burkhard Maess and Yvonne Wolff-Rosier for support and advice on all MEG-related aspects of this project. We thank Zejin Lu for supplying the ecoset-crop trained ResNet50 and AlexNet. This work was supported by the ERC Starting Grant TIME (101039524) awarded to T.C.K.; P.S. was supported by the Studienstiftung des deutschen Volkes, the Federal Ministry of Education and Research (BMBF), the Max Planck Society (MPG; Max-Planck-School of Cognition), and Deutsche Forschungsgemeinschaft (DFG; German Research Foundation) - GRK 2340. M.N.H. was supported by a research group grant by the Max Planck Society, the ERC Starting Grant COREDIM (101039712), a LOEWE Start Professorship of the Hessian Ministry of Higher Education, Science, Research and Art, and the Deutsche Forschungsgemeinschaft (DFG, German Research Foundation) under Germany's Excellence Strategy (EXC 3066/1 "The Adaptive Mind", Project No. 533717223). The Osnabrück HPC received support from the Deutsche Forschungsgemeinschaft (DFG, German Research Foundation) - 456666331.

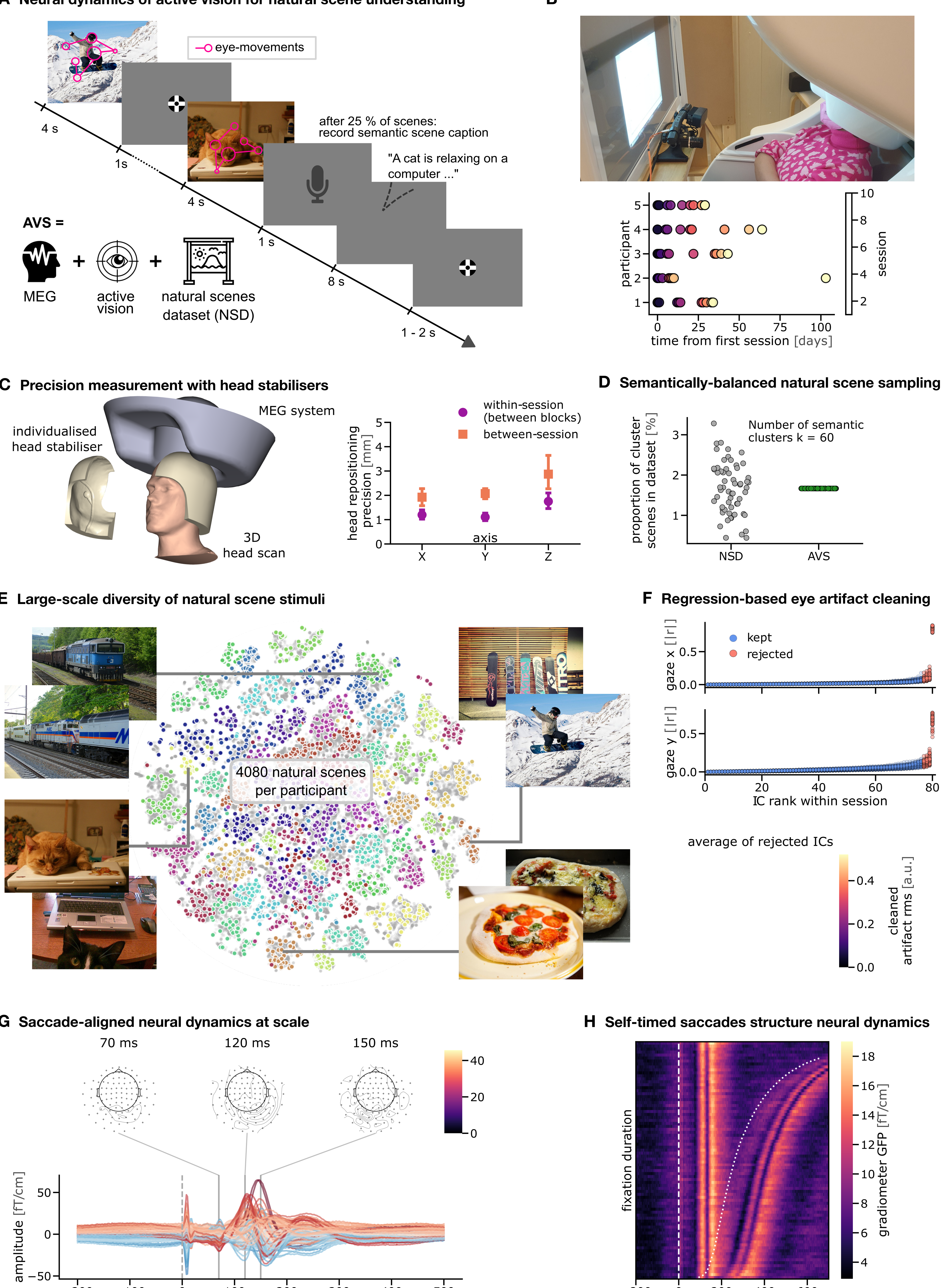
A Neural dynamics of active vision for natural scene understanding
eye-movements
4 s
1s
4 s
1 s
8 s
1 - 2 s
after 25 % of scenes:
record semantic scene caption
"A cat is relaxing on a computer ..."
AVS =
MEG
active vision
natural scenes dataset (NSD)
B
participant
session
time from first session [days]
C Precision measurement with head stabilisers
individualised head stabiliser
MEG system
3D head scan
head repositioning precision [mm]
within-session (between blocks)
between-session
axis
X
Y
Z
D Semantically-balanced natural scene sampling
proportion of cluster scenes in dataset [%]
Number of semantic clusters k = 60
NSD
AVS
E Large-scale diversity of natural scene stimuli
4080 natural scenes per participant
F Regression-based eye artifact cleaning
gaze x [|r|]
gaze y [|r|]
kept
rejected
IC rank within session
average of rejected ICs
cleaned artifact rms [a.u.]
G Saccade-aligned neural dynamics at scale
70 ms
120 ms
150 ms
amplitude [fT/cm]
time [ms]
H Self-timed saccades structure neural dynamics
fixation duration
gradiometer GFP [fT/cm]
time [ms]

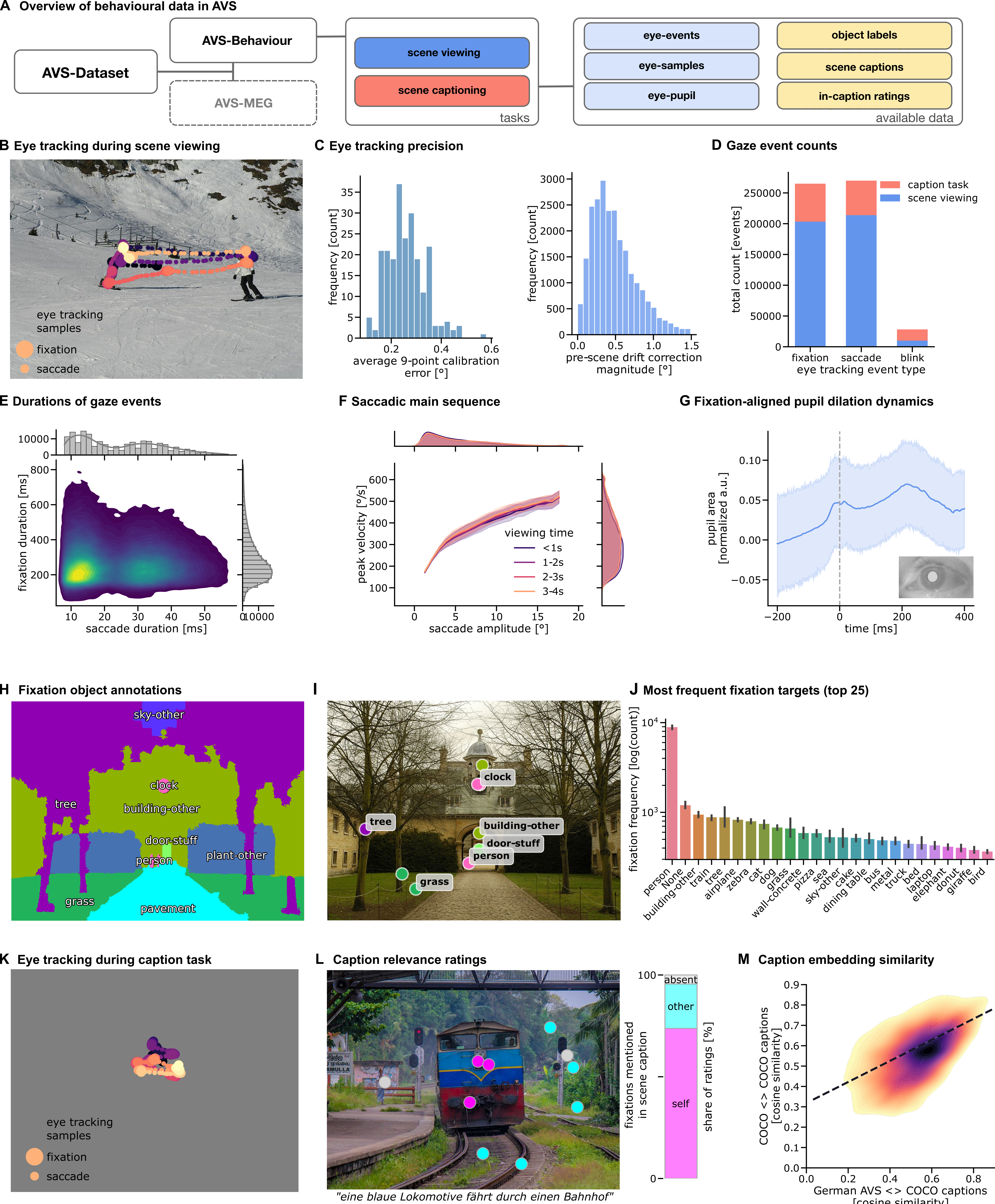
A Overview of behavioural data in AVS
AVS-Dataset
AVS-Behaviour
AVS-MEG
scene viewing
scene captioning
tasks
eye-events
eye-samples
eye-pupil
object labels
scene captions
in-caption ratings
available data
B Eye tracking during scene viewing
eye tracking samples
fixation
saccade
C Eye tracking precision
frequency [count]
average 9-point calibration error [°]
pre-scene drift correction magnitude [°]
D Gaze event counts
caption task
scene viewing
total count [events]
fixation
saccade
blink
eye tracking event type
E Durations of gaze events
fixation duration [ms]
saccade duration [ms]
F Saccadic main sequence
peak velocity [°/s]
viewing time
<1s
1-2s
2-3s
3-4s
saccade amplitude [°]
G Fixation-aligned pupil dilation dynamics
pupil area [normalized a.u.]
time [ms]
H Fixation object annotations
sky-other
clock
tree
building-other
door-stuff
person
plant-other
grass
pavement
I
clock
tree
building-other
door-stuff
person
grass
J Most frequent fixation targets (top 25)
fixation frequency [log(count)]
person
None
building-other
train
tree
airplane
zebra
cat
dog
grass
wall-concrete
pizza
sea
sky-other
cake
dining table
bus
metal
truck
bed
laptop
elephant
donut
giraffe
bird
K Eye tracking during caption task
eye tracking samples
fixation
saccade
L Caption relevance ratings
fixations mentioned in scene caption
absent
other
self
share of ratings [%]
"eine blaue Lokomotive fährt durch einen Bahnhof"
M Caption embedding similarity
COCO <> COCO captions [cosine similarity]
German AVS <> COCO captions [cosine similarity]

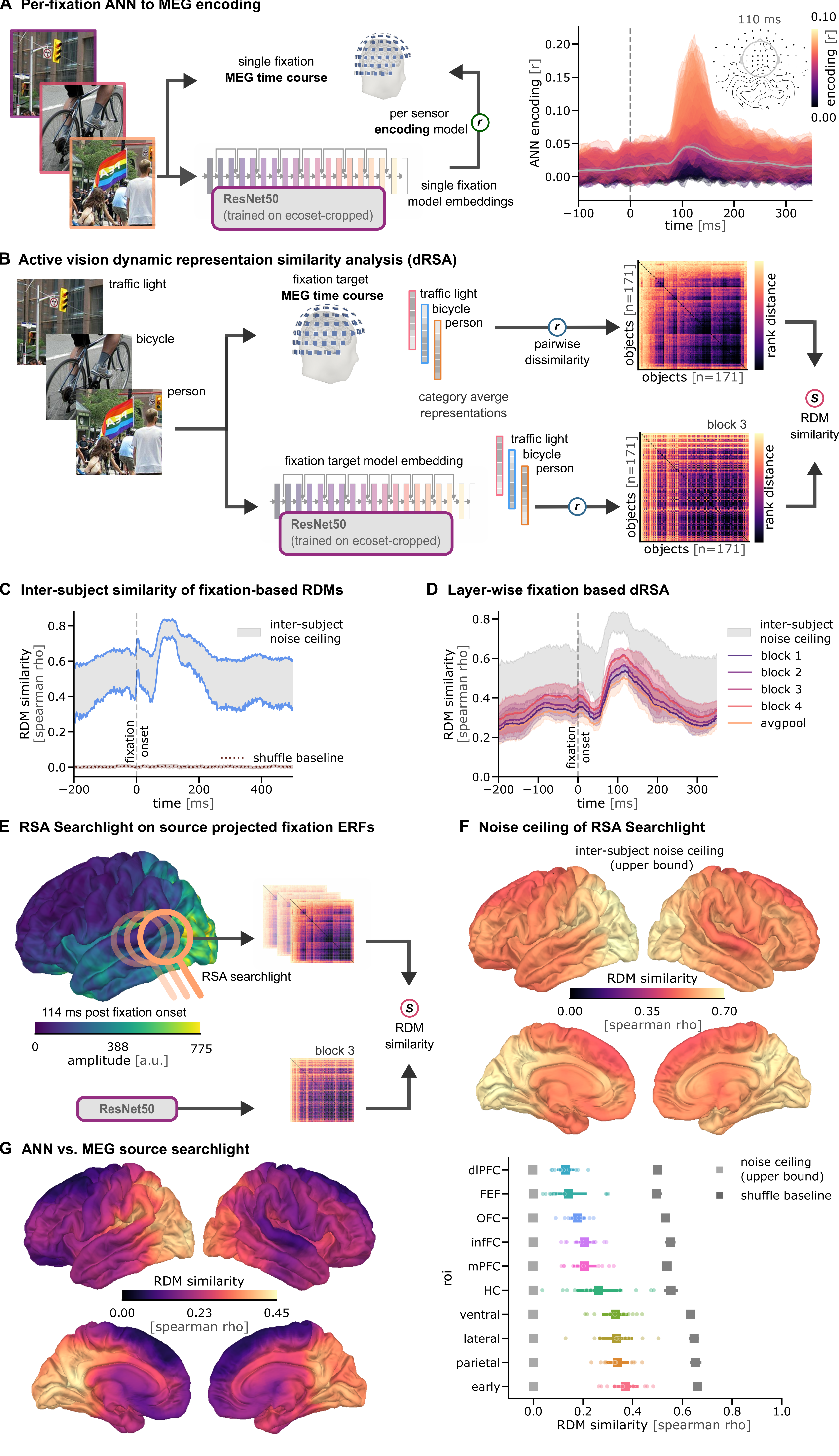

A Per-fixation ANN to MEG encoding
single fixation
MEG time course
per sensor
encoding model
ResNet50
(trained on ecoset-cropped)
single fixation
model embeddings
110 ms
ANN encoding [r]
encoding [r]
time [ms]
B Active vision dynamic representaion similarity analysis (dRSA)
traffic light
bicycle
person
fixation target
MEG time course
pairwise
dissimilarity
category averge
representations
objects [n=171]
rank distance
RDM
similarity
block 3
fixation target model embedding
C Inter-subject similarity of fixation-based RDMs
D Layer-wise fixation based dRSA
inter-subject
noise ceiling
shuffle baseline
fixation
onset
RDM similarity
[spearman rho]
block 1
block 2
block 3
block 4
avgpool
time [ms]
E RSA Searchlight on source projected fixation ERFs
F Noise ceiling of RSA Searchlight
inter-subject noise ceiling
(upper bound)
RSA searchlight
114 ms post fixation onset
amplitude [a.u.]
G ANN vs. MEG source searchlight
noise ceiling
(upper bound)
shuffle baseline
dlPFC
FEF
OFC
infFC
mPFC
HC
ventral
lateral
parietal
early
roi
RDM similarity [spearman rho]